\documentclass[11pt]{article}
\usepackage{latexsym,amssymb,amsmath}
\usepackage[dvips]{graphics,graphicx,epsfig,color}
\usepackage[lofdepth,lotdepth]{subfig}
\begin{document}
\title{\bf Generalization of Rayleigh's  high-frequency theory for the 2D Helmholtz equation in a half-space subject to a radiation condition at infinity and a Dirichlet condition on a 1D periodically-uneven boundary}
\author{Armand Wirgin\thanks{retired from LMA, CNRS, UPR 7051, Aix-Marseille Univ, Centrale
Marseille, F-13453 Marseille Cedex 13, France, ({\tt armand.wirgin@gmail.com})} }
\date{\today}
\maketitle
\begin{abstract}
The 2D Helmholtz equation, radiation condition and Dirichlet boundary condition, are the translation, in mathematical terms,  of (at least) three physical 2D  problems for the prediction of the total scalar wavefield on one side of an impenetrable 1D periodically uneven boundary  when: a) a plane TE electromagnetic wave propagating in the vacuum strikes the boundary the other side of which is occupied by a perfectly-conducting medium, b) a plane SH elastic elastic wave strikes a rigid boundary, c) a plane acoustic wave strikes a pressure-release boundary.
The first attempt to solve such problems in a non-heuristic manner was made by Lord Rayleigh (in
his book, 'The Theory of Sound'  which appeared in 1896). My task will be to revisit
Rayleigh's theory of diffraction by a sinusoidal-shaped, impenetrable boundary, and more specifically, his perturbation method for obtaining a
mathematically-explicit solution to the diffraction problem in the high-frequency regime. In so
doing, I shall correct and generalize Rayleigh's method to obtain  solutions for arbitrary angles of
incidence as well as for 1D periodic impenetrable boundaries of quite-general shape.
\end{abstract}
\newline
\newline
Abbreviated title: Rayleigh's diffraction grating theory for a Dirichlet boundary condition revisited
\newline
\newline
Corresponding author: Armand Wirgin, \\ e-mail: armand.wirgin@gmail.com
\newpage
\tableofcontents
\newpage
\newpage
\section{Introduction}\label{intro}
\subsection{Rayleigh's contribution}\label{raycontrib}
Perhaps the best introduction to the present study is to cite none other than Rayleigh's
contribution, in sect 272a of his book \cite{ra96}, in which the wavefield refers to the acoustic velocity potential.\\

'In the preceding sections the surface of separation at which reflection takes place, is
supposed to be absolutely plane. It is of interest, both from an acoustical and from an optical point of view, to inquire what effect would be produced by roughnesses, or corrugations, in the reflecting surface; and the problem thus presented may be solved without difficulty to a
certain extent by the method of \S268, especially if we limit ourselves to the case of
perpendicular (I call this normal) incidence. The equation of the reflecting surface will be
supposed to be $z=\zeta$ where $\zeta$ is a periodic function (I call $d$ the period) of $x$
whose mean value is zero. As a particular case we may take
\begin{equation}\label{ray_1}
    \zeta=c~\cos px
\end{equation}
(I denote his $c$ by $h$, and his $p$ by $2\pi/d$), but in general we should have to supplement the
first term of the series expressed in (\ref{ray_1}) by cosines and sines of the multiples of
$px$. The velocity potential of the incident wave (of amplitude unity) may be written
\begin{equation}\label{ray_2}
\phi=e^{\imath k(at+z)}~,
\end{equation}
(I, as well as Rayleigh, denote $k$ and $t$ by the wavenumber and temporal variable
respectively, and his $a$ is the wavespeed which I designate by $-c$ ).

For the regularly reflected wave we have $\phi=A_{0}e^{\imath kz}$, the time factor being dropped
for the sake of brevity; but to this must be added terms in $\cos px$, $\cos 2px$, etc. Thus,
as the complete value of $\phi$ in the upper medium,
\begin{equation}\label{ray_3}
    \phi=e^{\imath kz}+A_{0}e^{-\imath kz}+A_{1}e^{-\imath\mu_{1})}\cos px+A_{2}e^{-\imath\mu_{2}z}\cos 2px+...
\end{equation}
in which
\begin{equation}\label{ray_4}
    \mu_{1}^{2}=k^{2}-p^{2}~,~\mu_{2}^{2}=k^{2}-4p^{2},...
\end{equation}

The expression (\ref{ray_3}), in which for simplicity sines of multiples of $px$ have been
omitted from the first, would be sufficiently general even though cosines of multiples of $px$
accompanied $c \cos px$ in (\ref{ray_1}).'\\

Further on in sect 272a, Rayleigh writes\\

'The full solution of the problem of the present section would require the determination of the
reflection  when $k$ is given for all values of $c$ and for all values of $p$.....we shall
presently deal with the case where $p<k$ (it would be more correct to write $p<<k$). For intermediate
values of $p$ the problem is more difficult, and in considering them we shall limit ourselves
to the simpler boundary conditions which obtain when no energy penetrates the second medium .
The simplest case of all (i.e., the third configuration of my abstract and such
that).... the boundary condition (7) reduces to
\begin{equation}\label{ray_15}
    \phi=0~.
\end{equation}
By (\ref{ray_3}) and (\ref{ray_15}) ((15) in Rayleigh's book), the condition to be satisfied at the surface
is
\begin{equation}\label{ray_16}
    e^{2\imath kz}+A_{0}+A_{1}e^{\imath (k-\mu_{1})z}\cos px+A_{2}e^{\imath (k-\mu_{2})z}\cos 2px+...=0~.
\end{equation}
.\\
.\\
.\\\\
In our problem $z$ is given by (\ref{ray_1}) as a function of $x$ and the equations of
condition are to be found by equating to zero the coefficients of the various terms involving
$\cos px$, $\cos 2px$, etc., when the left hand member of (16) (here (\ref{ray_16})) is
expanded in Fourier's series. The development of the various exponentials is effected as in
(12), and the resulting equations are
\begin{equation}\label{ray_17}
  J_{0}(2k)+A_{0}+\imath A_{1}J_{1}(k-\mu_{1})-A_{2}J_{1}(k-\mu_{2})-... =0
\end{equation}
\begin{equation}\label{ray_18}
  2\imath J_{1}(2k)+A_{1}[J_{0}(k-\mu_{1})-J_{2}(k-\mu_{1})]+A_{2}[\imath J_{1}(k-\mu_{2})-i
  J_{3}(k-\mu_{2})]+... =0
\end{equation}
\begin{equation}\label{ray_19}
  -2J_{2}(2k)+A_{1}[i J_{1}(k-\mu_{1})-\imath
  J_{3}(k-\mu_{1})]+A_{2}[J_{0}(k-\mu_{2})+J_{4}(k-\mu_{2})]+... =0
\end{equation}
and so on, where for the sake of brevity $c$ has been made equal to unity. So far as
$(k-\mu)$ may be treated as real, as happens for a large number of terms when $p$ is small
relatively to $k$, the various Bessel's functions are all real and thus the $A$'s of even order
are real and the $A$'s of odd order are pure imaginaries. Accordingly, the phase of the
perpendicularly reflected wave is the same as if $c=0$; but it must be remembered that this
conclusion is in reality only approximate, because, however small $p$ may be, the $\mu$'s end
by becoming imaginary.\\

From the above equations it is easy to obtain the value of $A_{0}$ as far as the term in
$p^{4}$. From (\ref{ray_19})((19) in Rayleigh's book)
\begin{equation}\label{ray_20a}
    A_{2}=2J_{2}(2k)~;
\end{equation}
from (\ref{ray_18})~ ((18) in Rayleigh's book)
\begin{equation}\label{ray_20b}
    \imath A_{1}=2J_{1}(2k)+(k-\mu_{2})J_{2}(2k)~;
\end{equation}
and finally from (\ref{ray_17}) ((17) in Rayleigh's book)
\begin{equation}\label{ray_20c}
     -A_{0}=J_{0}(2k)+(k-\mu_{1})J_{1}(2k)+\Big\{\frac{1}{2}(k-\mu_{1}))(k-\mu_{2})-
     \frac{1}{4}(k-\mu_{2})^{2}\Big\}J_{2}(2k)+...~.
\end{equation}
\\
From (\ref{ray_4})
\begin{equation}\label{ray_21a}
     k-\mu_{1}=\frac{p^{2}}{2k}+\frac{p^{4}}{8k^{3}}+...~;
\end{equation}
so that, as expanded in powers of $p$ with reintroduction of $c$,
\begin{equation}\label{ray_21b}
     -A_{0}=J_{0}(2kc)+\frac{p^{2}}{k^{2}}\frac{1}{2}kc
     J_{1}(2kc)+\frac{p^{4}}{k^{4}}\Big\{\frac{1}{8}kc
     J_{1}(2kc)-\frac{1}{2}k^{2}c^{2}J_{2}(2kc)\Big\}+...~.
\end{equation}
This gives the amplitude of the perpendicularly reflected wave, with omission of $p^{6}$ and
higher orders of $p$.'\\
.\\
.\\
.\\
\subsection{Comments on Rayleigh's contribution and the need for its generalization}
In his book 'The Theory of Sound', Rayleigh contributes two essential features to the
mathematical resolution of the 2D problem of the diffraction of a plane wave by a
1D periodically-uneven opaque boundary: a) a representation ((\ref{ray_3})), at \emph{all
locations} (including the boundary), of the diffracted wavefield as a discrete superposition of
propagative and evanescent plane waves whose amplitudes do not depend on the space variables,
and b) a perturbation method, for high temporal frequencies, to solve, in
mathematically-explicit manner, for the plane wave amplitudes.

I shall not comment herein on whether the Rayleigh plane wave representation is valid, not
valid, or otherwise, since this issue has been abundantly discussed in many other publications \cite{bs63},\cite{wi80},\cite{wi82}.
Thus, I will assume that Rayleigh's representation is valid. As for his perturbation method,
other than the fact that Rayleigh's treatment is extremely concise and, in his own words
'easy', it is sometimes error-ridden and suffers from two major limitations which are that it has been applied: 1) only to
the case of a sinusoidally-uneven boundary, and 2) and only to the case of a
perpendicularly-incident (plane) wave. The first limitation is annoying since, in many
practical applications (e.g., optical spectroscopy), diffraction gratings have periodic profiles that
are not sinusoidal. The second limitation is troublesome because it is impossible to
accurately-measure the backscattered energy (and thus $|A_{0}|$) in any attempt to
experimentally-verify the Rayleigh solutions) for perpendicular incidence. Consequently, I
shall not limit my treatment of this diffraction problem to sinusoidal boundaries nor to
perpendicularly-incident plane waves, but the price for doing so is a certain amount of
complexity. Also, to conform with my earlier treatments of the diffraction grating problem, I
shall employ symbols for the configuration constants and field variables that are different
from those of Rayleigh, which fact should not be source of difficulties since my treatment of
this problem is self-contained (i.e., does not rely on any of Rayleigh's formulae).
\section{The boundary value problem}\label{boundval}
\subsection{The scattering configuration}\label{scatconfig}
\begin{figure}[ht]
\begin{center}
\includegraphics[width=0.75\textwidth]{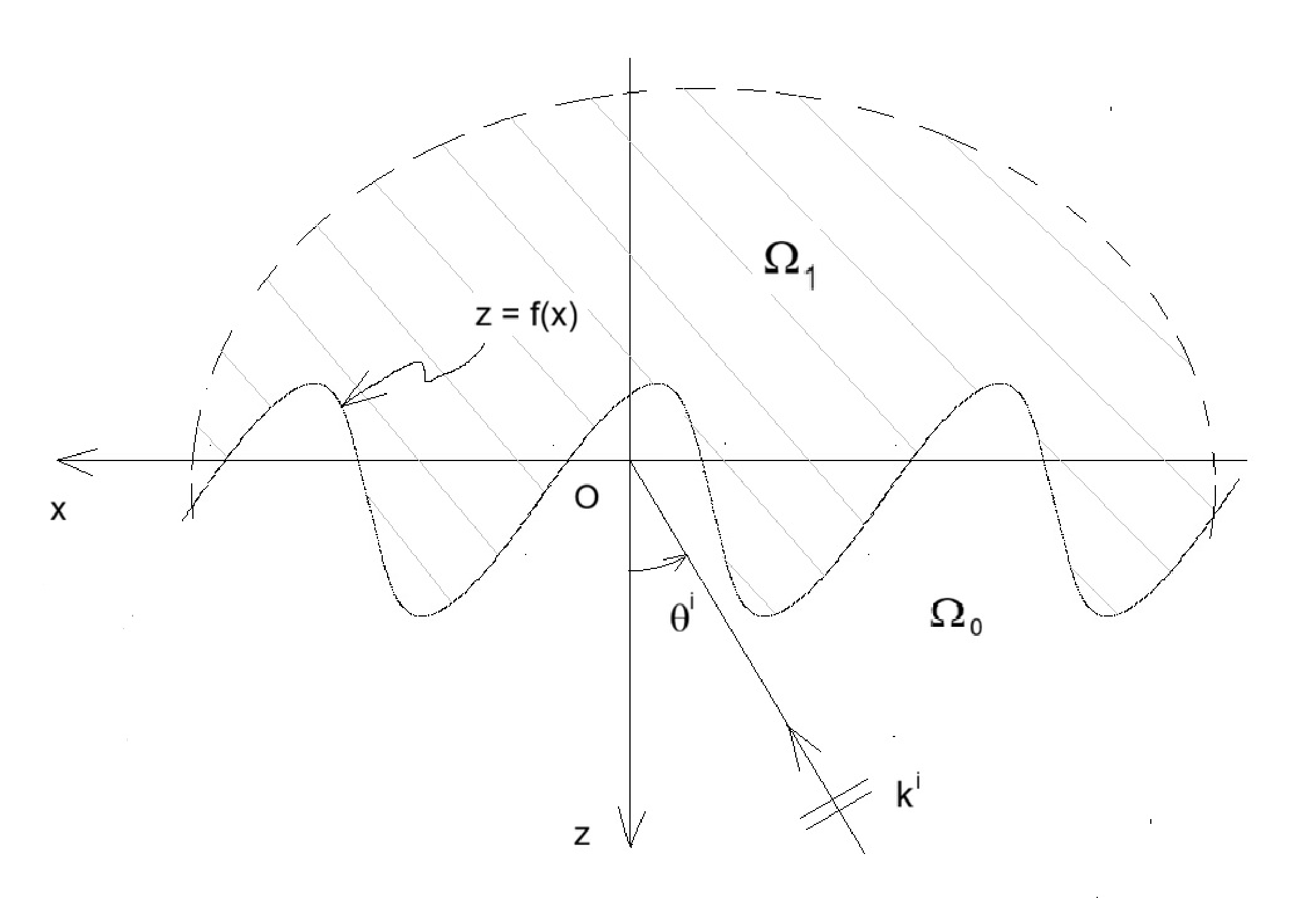}
\caption{Sagittal plane view of 2D scattering configuration.}
\label{config}
\end{center}
\end{figure}

In what follows, I adopt the language   of the scattering (i.e., diffraction) of a
monochromatic electromagnetic (e.g., optical) wave by an impenetrable boundary.

Fig. \ref{config} describes the scattering configuration in which two media are separated
by a periodically-uneven (i.e., non-planar)  boundary. The bottom medium is the vacuum in which
propagates the plane wave field, whereas the top medium is perfectly-conducting, which means
that the electromagnetic field is nil therein (and therefore of no interest). In the figure,
in which $Oxyz$ is the cartesian coordinate system, the periodically-uneven (i.e., non-planar)
interface is described by the equation $z=f(x)=f(x+d): \forall x\in \mathbb{R},\forall y\in \mathbb{R}$, $d$ is
the geometric period of the boundary,
$\mathbf{k}^{i}=\mathbf{k}^{i}(\theta^{i},\omega)=(k_{x}^{i},0,-k_{z}^{i})=(k s^{i},0,kc^{i})$
is the wavevector of the incident plane wave, oriented so that its $y$  component is nil,
$\theta^{i}$  the angle of incidence, $s^{i}=\sin\theta^{i}$, $c^{i}=\cos\theta^{i}$,
$\|k^{i}\|=k=\omega/c$, with $c$ the velocity of electromagnetic waves in the vacuum, and
$\omega$ the angular frequency. The incident vectorial wavefield is assumed to be
transverse-electric (TE)-polarized which means that the incident electric field (in the bottom
medium) is of the form $\mathbf{E}^{i}=(0,E^{i}_{y}(x,z,t),0)$ with $t$ the time). Since
neither the scattering boundary nor the incident wavefield depend on $y$, it ensues that the
scattered and total wavefields in the bottom medium do not depend on $y$ either.  Thus, the
problem is 2D  and all the fields (whose $\exp(-\imath\omega t)$ temporal dependence is henceforth
implicit) are describable in the sole $xOz$ (sagittal) plane wherein I define $\Omega_{0}$ and
$\Omega_{1}$ to be the bottom and top half-planes respectively on the two sides of  the
non-linear curve $z=f(x)$.
\subsection{Mathematical formulation of the boundary-value problem}
With the definitions: $\mathbf{x}=(x,z)$ and $u^{i}(\mathbf{x})=E^{i}_{y}(\mathbf{x})$, the
to-be-determined scattered and total wavefields take the forms
$u^{s}(\mathbf{x})=E^{s}_{y}(\mathbf{x })$ and $u(\mathbf{x})=E_{y}(\mathbf{x})$ respectively,
with the understanding that
\begin{equation}\label{ray_30}
 u(\mathbf{x})=u^{i}(\mathbf{x})+u^{s}(\mathbf{x})~
\end{equation}
and
\begin{equation}\label{ray_40}
 u^{i}(\mathbf{x})=\exp(\imath \mathbf{k}^{i}\cdot\mathbf{x})~.
\end{equation}
On account of the independence of the wavefield on $y$, the TE polarization and the
impenetrable nature of the rough boundary, Maxwells' equations reduce to the single 2D partial
differential Helmholtz equation
\begin{equation}\label{ray_50}
 \big[u_{,xx}(\mathbf{x})+u_{,zz}(\mathbf{x})+k^{2}u(\mathbf{x})\big]=0~;\forall
 \mathbf{x}\in\Omega_{0}~,
\end{equation}
(wherein $F_{,\eta}=\frac{\partial F}{\partial\eta}$ and
$F_{,\eta\eta}=\frac{\partial^{2}F}{\partial \eta^{2}}$) and to the Dirichlet boundary condition
\begin{equation}\label{ray_60}
 u(x,f(x))=0~;~\forall x\in \mathbb{R}~.
\end{equation}
Finally, the scattered wavefield must satisfy the radiation condition
\begin{equation}\label{ray_70}
 u^{s}(\mathbf{x})\sim \text{outgoing waves}~;~\|\mathbf{x}\|\rightarrow\infty~,
\end{equation}
with the understanding that $u^{i}$ is an incoming wave with respect to the scattering
boundary.
\section{Wavefield representation}\label{urep}
Eq. (\ref{ray_40}) implies that
\begin{equation}\label{ray_80}
 u^{i}(x+d,z)=u^{i}(x,z)\exp(\imath ks^{i}d)~;~\forall \mathbf{x}\in \Omega_{0}~,
\end{equation}
whence, on account of (\ref{ray_50}), (\ref{ray_60}) and the $d$-periodicity of $f$,
\begin{equation}\label{ray_90}
 u(x+d,z)=u(x,z)\exp(\imath ks^{i}d)~;~\forall \mathbf{x}\in \Omega_{0}~,
\end{equation}
so that separation of variables as well as the radiation condition translate to the following
representation of $u$:
\begin{equation}\label{ray_100}
 u(x,z)=u^{i}(x,z)+\sum_{n\in\mathbb{Z}}B_{n}\exp[\imath(k_{xn}x+k_{zn}z)]~;~\forall \mathbf{x}\in
 \Omega_{0}~,
\end{equation}
in which
\begin{equation}\label{ray_110}
 k_{xn}=k_{x}^{i}+\frac{2\pi n}{d}~,~k_{zn}=[1-(k_{xn}^2)]^{\frac{1}{2}}~,\Re(k_{zn})\ge
 0~,~\Im(k_{zn})\ge 0~,
\end{equation}
or using a notation closer to that of Rayleigh,
\begin{equation}\label{ray_120}
 u(x,z)=\exp[\imath k(xs^{i}-zc^{i})]+
 \sum_{n=-\infty}^{\infty}B_{n}\exp[\imath k(xs_{n}x+zc_{n})]~;~\forall \mathbf{x}\in \Omega_{0}~,
\end{equation}
in which
\begin{equation}\label{ray_130}
 s_{n}=s^{i}+\frac{2\pi n}{kd}~,~c_{n}=[1-(s_{n}^2)]^{\frac{1}{2}}~,\Re(c_{n})\ge
 0~,~\Im(c_{n})\ge 0~.
\end{equation}
For normal (i.e., perpendicular) incidence, $\theta^{i}=0$, $s^{i}=0$ and $c^{i}=1$, so that
$s_{-n}=-s_{n}$ and $c_{-n}=c_{n}$ whence a straightforward consequence of (\ref{ray_120}) is
that
\begin{equation}\label{ray_140}
 u(-x,z)=u(x,z)~;~\forall \mathbf{x}\in \Omega_{0}~,
\end{equation}
the result of which is that $B_{-n}=B_{n}$ so that it is easy to show (from  (\ref{ray_120}))
that
\begin{multline}\label{ray_150}
 U(x,z)=u(x,z)\exp[-\imath k(xs^{i}+zc^{i})]=\\
 \exp(-\imath 2kzc^{i})+A_{0}\exp(\imath 2kzc^{i})+
 \sum_{n=1}^{\infty}A_{n}\cos\frac{2\pi n}{d}x\exp[\imath k(c_{n}-c_{0})z]~;~\forall \mathbf{x}\in\Omega_{0}~,
\end{multline}
(wherein $A_{0}=B_{0}$ and $A_{n>0}=2B_{n>1}$),
 which explains the origin  of Rayleigh's equation (\ref{ray_3}) (also stemming from the
 assumption of normal incidence). \emph{For other-than-normal incidence, (\ref{ray_150}) is not
 a complete representation and must be replaced by the plane-wave-representation
 (\ref{ray_120})}.
\section{Application of the plane-wave-representation in the Dirichlet boundary condition to obtain the $B_{n}$}
From (\ref{ray_60}) and (\ref{ray_120}) it follows that
\begin{equation}\label{ray_160}
 u(x,f(x))=0=\exp[\imath k(xs^{i}-f(x)c^{i})]+
 \sum_{n=-\infty}^{\infty}B_{n}\exp[\imath k(s_{n}x+f(x)c_{n})]~;~\forall x\in [0,d]~,
\end{equation}
which, after  projection, yields
\begin{multline}\label{ray_170}
 \int_{0}^{d}u(x,f(x))\exp([-\imath k(s_{m}x+c_{0}f(x))]dx=\\
 0=\int_{0}^{d}\exp[\imath k(x(s^{i}-s_{m})-2f(x)c^{i})]dx+
 \sum_{n\in\mathbb{Z}}B_{n}\int_{0}^{d}\exp[
 \imath k(x(s_{n}-s_{m})+f(x)(c_{n}-c_{0}))]dx~;~\forall
 m\in \mathbb{Z}~,
\end{multline}
or, after the change of variables $x=\zeta d$, $f(x)=f(\zeta d)=g(\zeta)$,
\begin{equation}\label{ray_180}
 \sum_{n=-\infty}^{\infty}\gamma_{mn}B_{n}=\alpha_{m}~;~\forall m\in\mathbb{Z}~,
\end{equation}
in which:
\begin{multline}\label{ray_190}
 \gamma_{mn}=\int_{0}^{1}\exp[\imath(2\pi(n-m)\zeta+(c_{n}-c_{0})kg(\zeta))]d\zeta~,~\\
 \alpha_{m}=-\int_{0}^{1}\exp[\imath(-2\pi m\zeta-2c_{0}kg(\zeta))]d\zeta
 ~.
\end{multline}
Eq. (\ref{ray_180}) is the infinite set of linear equations by which the infinite set of
unknown plane wave amplitudes ($B_{n}$) can, in principle, be numerically solved-for. I refer
the interested reader to the literature \cite{bs63},\cite{wi80}, since this task is not
the object of the present
study. Rather, I shall now expose the high-frequency perturbation scheme by which I obtain
mathematically-explicit solutions for the $B_{n}$ starting from (\ref{ray_180})
\section{The perturbation scheme for 'arbitrary' periodic boundary profiles}
Henceforth, contrary to what Rayleigh assumed, I shall place no restrictions on the periodic
boundary profile other than
\begin{equation}\label{ray_195}
\text {$g(\zeta)$, and some of, or all, its derivatives are  bounded~, continuous  functions of $\zeta$}.
\end{equation}

Let
\begin{equation}\label{ray_200}
 \epsilon=\frac{2\pi}{kd}=\frac{2\pi c}{\omega d}~
\end{equation}
It can be observed, in (\ref{ray_180}), that the sole variable which depends on $\epsilon$ is
 $c_{n}=[1-(s^{i}+n\epsilon)]^{\frac{1}{2}}$.

High frequencies correspond to large $\omega$, or more specifically, to $kd >>2\pi$, which is
equivalent to $\epsilon << 1$ (i.e., small $\epsilon$). I assume that this is actually the
case. Then, it is appropriate to expand all three quantities in (\ref{ray_180}) in Taylor
series
\begin{equation}\label{ray_210}
 \gamma_{mn}=\sum_{q=0}^{\infty}\gamma_{mn}^{(q)}\epsilon^{q}~,~
 B_{n}=\sum_{p=0}^{\infty}B_{n}^{(p)}\epsilon^{p}~,~
 \alpha_{m}=\sum_{l=0}^{\infty}\alpha_{m}^{(l)}\epsilon^{l}~,
\end{equation}
wherein
\begin{equation}\label{ray_220}
\gamma_{mn}^{(q)}=\frac{1}{q!}\frac{\partial^{q}}{\partial\epsilon^{q}}\gamma_{mn}
\Big|_{\epsilon=0}~,~
\alpha_{m}^{(l)}=\frac{1}{l!}\frac{\partial^{l}}{\partial\epsilon^{l}}\alpha_{mn}
\Big|_{\epsilon=0}~.
\end{equation}
The introduction of (\ref{ray_210}) into (\ref{ray_180}) gives rise, after comparison of
powers of $\epsilon$, to
\begin{equation}\label{ray_230}
 \sum_{p=0}^{l}\sum_{n=-\infty}^{\infty}\gamma_{mn}^{(l-p)}B_{n}^{p}=\alpha_{m}^{(l)}~;
 ~l=0,1,2,...~,
\end{equation}
and since
\begin{equation}\label{ray_240}
 \gamma_{mn}^{(0)}=\int_{0}^{1}\exp[i2\pi(n-m)\zeta]d\zeta=\delta_{mn}~,
\end{equation}
wherein $\delta_{mn}$ is the Kronecker delta symbol, it follows that
\begin{equation}\label{ray_250}
 B_{m}^{(l)}=\alpha_{m}^{(l)}-\sum_{p=0}^{l-1}\sum_{n=-\infty}^{\infty}\gamma_{mn}^{(l-p)}B_{n}^{p}~;~l=0,1,2,...~,
\end{equation}
which is a recursion relation for the $B_{m}^{(l)}$. Eq. (\ref{ray_250}) is the principal
result of the perturbation scheme. I now show how to obtain the explicit solutions for
$l=0,1,2....$ (actually, due to the complexity of the involved algebra, I shall consider only
$l=0,1,2$). The result of all this will be explicit mathematical expressions for the zeroth,
first and second $\epsilon$-perturbation approximations of $B_{m}$, i.e.,
\begin{equation}\label{ray_255}
 _{0}B_{m}^{(l)}=B_{m}^{(0)}~,~_{1}B_{m}=B_{m}^{0}+\epsilon B_{m}^{1}~,~
 _{2}B_{m}=B_{m}^{0}+\epsilon B_{m}^{1}+\epsilon^{2}B_{m}^{(2)}~.
\end{equation}
\subsection{The $l=0$ perturbation order}\label{leq0}
The $l=0$ solution is $B_{m}^{(0)}=\alpha_{m}^{(0)}$, or
\begin{equation}\label{ray_260}
    B_{m}^{(0)}=-\int_{0}^{1}\exp[i(-2\pi m\zeta-2c_{0}kg)]d\zeta~,
\end{equation}
in which I have adopted (and will henceforth employ) the short-hand notation $g$ for
$g(\zeta)$.
\subsection{The $l=1$ perturbation order}\label{leq1=}
Since $\alpha_{m}$ depends only on $c_{0}$ which is a constant with respect to $\epsilon$,
\begin{equation}\label{ray_270}
    \alpha_{m}^{(l>0)}=0~,
\end{equation}
whence
\begin{equation}\label{ray_280}
    B_{m}^{(1)}=-\sum_{n=-\infty}^{\infty}\gamma_{mn}^{(1)}B_{n}^{(0)}~.
\end{equation}
From the fact that $(c_{n})_{,\epsilon}|_{\epsilon=0}=-\frac{ns_{0}}{c_{0}}$ I find %
\begin{equation}\label{ray_290}
    \gamma_{mn}^{(1)}=-ik\frac{ns_{0}}{c_{0}}\int_{0}^{1}g \exp[i(2\pi(n-m)\zeta)]d\zeta~,
\end{equation}
so that
\begin{equation}\label{ray_300}
    B_{m}^{(1)}=-\imath k\frac{s_{0}}{c_{0}}\int_{0}^{1}d\zeta' g' \exp[-\imath 2\pi m\zeta')]
    \int_{0}^{1}d\zeta\exp[-\imath 2c_{0}kg]\sum_{n=-\infty}^{\infty} n\exp[\imath 2\pi n(\zeta'-\zeta)]~.
\end{equation}
From the Poisson sum formula \cite{mf53}, I find
\begin{equation}\label{ray_310}
    \sum_{n=-\infty}^{\infty} \exp[\imath 2\pi
    n(\zeta'-\zeta)]=\delta(\zeta'-\zeta)~;~\zeta',\zeta\in[0,1]~,
\end{equation}
wherein $\delta(\zeta'-\zeta)$ is the Dirac delta distribution. It follows that
\begin{equation}\label{ray_320}
    \sum_{n=-\infty}^{\infty} n\exp[\imath 2\pi
    n(\zeta'-\zeta)]=\frac{\imath}{2\pi}[\delta(\zeta'-\zeta)]_{,\zeta}~;~\zeta',\zeta\in[0,1]~,
\end{equation}
whereas integration by parts  and the periodicity of $g$(i.e., $g(0)=g(1)$ entail %
\begin{equation}\label{ray_330}
     \int_{0}^{1}d\zeta\exp(-\imath 2c_{0}kg)[\delta(\zeta'-\zeta)]_{,\zeta}=\imath 2c_{0}k\int_{0}^{1}d\zeta
     g_{,\zeta}\exp(-\imath 2kc_{0}g)~,
\end{equation}
whence finally
\begin{equation}\label{ray_340}
     B_{m}^{(1)}=\frac{\imath k^{2}s_{0}}{\pi}\int_{0}^{1}gg_{,\zeta}\exp[-\imath(2\pi
     m\zeta+2c_{0}kg)]d\zeta~.
\end{equation}
\subsection{The $l=2$ perturbation order}\label{leq2}
The point of departure is
\begin{equation}\label{ray_350}
     B_{m}^{(2)}=-\sum_{n=-\infty}^{\infty}\gamma_{mn}^{(2)}B_{n}^{(0)}-
     \sum_{n=-\infty}^{\infty}\gamma_{mn}^{(1)}B_{n}^{(1)}=\Sigma_{20}+\Sigma_{11}~.
\end{equation}
Using the techniques employed for $l=1$, I find: %
\begin{equation}\label{ray_360}
     \gamma_{mn}^{(2)}=-\frac{n^{2}}{2}\int_{0}^{1}\Big[\frac{\imath k}{c_{0}}
     \Big(1+\frac{s_{0}^{2}}{c_{0}^{2}}\Big)g+
     k^{2}\frac{s_{0}^{2}}{c_{0}^{2}}g^{2}\Big]\exp[\imath 2\pi(n-m)\zeta]d\zeta~,
\end{equation}
\begin{equation}\label{ray_370}
     \Sigma_{20}=-\Big(\frac{\imath 2kc_{0}}{8\pi^{2}}\Big)\int_{0}^{1}\Big[\frac{ik}{c_{0}}
     \Big(1+\frac{s_{0}^{2}}{c_{0}^{2}}\Big)g+
     k^{2}\frac{s_{0}^{2}}{c_{0}^{2}}g^{2}\Big]\Big[g_{,\zeta\zeta}-\imath 2kc_{0}(g_{,\zeta})^2\Big]
     \exp[-i
     \imath(2\pi m\zeta+2c_{0}kg)]d\zeta~,
\end{equation}
\begin{equation}\label{ray_380}
     \Sigma_{11}=\Big(\frac{\imath k^{3}s_{0}^{2}}{2\pi^{2}c_{0}}\Big)
     \int_{0}^{1}g\Big[(g_{,\zeta})^{2}+gg_{,\zeta\zeta}-\imath 2kc_{0}g(g_{,\zeta})^{2}\Big]
     \exp[-\imath(2\pi m\zeta+2c_{0}kg)]d\zeta~,
\end{equation}
so that finally
\begin{multline}\label{ray_390}
     B_{m}^{(2)}=\Big(\frac{k^{2}}{4\pi^{2}}\Big)
     \int_{0}^{1}g\Big[\Big(\frac{1}{c_{0}^{2}}g g_{,\zeta\zeta}-\imath 2kc_{0}g(g_{,\zeta})^{2}+
     \Big(\frac{\imath ks_{0}^{2}}{c_{0}}\Big)g^{2}g_{,\zeta\zeta}+
     2k^{2}s_{0}^{2}g^{2}(g_{,\zeta})^{2}\Big]\times\\
     \exp[-\imath(2\pi m\zeta+2c_{0}kg)]d\zeta~,
\end{multline}
\section{$\epsilon$-perturbation approximations for a sinusoidal boundary struck by a plane
wave of arbitrary angle of incidence}
Until now, the periodic boundary profile function $f(x)$, and therefore $g(\zeta)$, were
assumed to be arbitrary (although bounded and twice differentiable) and $\theta^{i}$ was
assumed to be arbitrary. Now, I again make the second assumption and restrict $f$ to the case
(the only one studied by Rayleigh) in which it is a sinusoidal function:
\begin{equation}\label{ray_400}
    f(x)=h\cos \Big(\frac{2\pi}{d}x\Big)~~\Rightarrow~~g(\zeta)=h\cos(2\pi\zeta)
\end{equation}
\subsection{The $l=0$ approximation}
I make use of the well-known integral definition of the $m$-th order Bessel function of the
first kind \cite{as68}:
\begin{equation}\label{ray_410}
    \int_{0}^{1}\exp[-\imath(2\pi m\zeta+\eta\cos 2\pi\zeta)]d\zeta=(-\imath)^{m}J_{m}(\eta)~
\end{equation}
whence
\begin{equation}\label{ray_420}
    _{0}B_{m}=B_{m}^{(0)}=-(-\imath)^{m}J_{m}(2khc_{0})~.
\end{equation}
\subsection{The $l=1$ approximation}
The fact that $g_{,\epsilon}=-2\pi h\sin 2\pi\zeta$ entails $gg_{,\eta}=\frac{1}{2}\sin
4\pi\zeta=\frac{1}{4}[\exp (i4\pi\zeta)-\exp (-i4\pi\zeta)]$, so that via (\ref{ray_410}),
(\ref{ray_340}) becomes
\begin{equation}\label{ray_430}
    B_{m}^{(1)}=(kh)^{2}\frac{s_{0}}{2}(-\imath)^{m}
    [J_{m-2}(2khc_{0})-J_{m+2}(2khc_{0})]~
\end{equation}
the consequence of which is
\begin{equation}\label{ray_440}
    _{1}B_{m}=-(-\imath)^{m}J_{m}(2khc_{0})+\epsilon(kh)^{2}\frac{s_{0}}{2}(-i)^{m}
    [J_{m-2}(2khc_{0})-J_{m+2}(2khc_{0})]~.
\end{equation}
\subsection{The $l=2$ approximation}
In the same manner, I find
\begin{multline}\label{ray_450}
    B_{m}^{(2)}=-\frac{(k h)^{2}}{4c_{0}}(-\imath)^{m}
    \Big[\frac{1}{c_{0}}[2J_{m}(2khc_{0})-J_{m-2}(2khc_{0})-J_{m+2}(2khc_{0})]-\\
    k h[J_{m-1}(2khc_{0})-J_{m+1}(2khc_{0})+J_{m-3}(2khc_{0})-J_{m+3}(2khc_{0})]+\\
    \frac{k h
    s_{0}^{2}}{2}[-J_{m-1}(2khc_{0})+J_{m+1}(2khc_{0})+3J_{m-3}(2khc_{0})-3J_{m+3}(2khc_{0})]-\\
    \frac{(k h)^{2}s_{0}^{2}c_{0}}{2}[2J_{m}(2khc_{0})-J_{m-4}(2khc_{0})-J_{m+4}(2khc_{0})]\Big]~
\end{multline}
which, by making use of the identity \cite{as68}
$J_{n-1}(\eta)+J_{(n+1)}(\eta)=\frac{2n}{\eta}J_{n}(\eta)$ for $n=m\pm 1$, $n=m\pm 2$ and
$n=m\pm 3$, reduces to
\begin{multline}\label{ray_460}
    B_{m}^{(2)}=-\frac{k h}{4c_{0}^{3}}(-\imath)^{m}\Big[[J_{m+1}(2khc_{0})-J_{m-1}(2khc_{0})]+\\
    m[J_{m-1}(2khc_{0})+J_{m+1}(2khc_{0})]
    -mk h c_{0}[J_{m-2}(2khc_{0})-J_{m+2}(2khc_{0})]-\\
m\frac{(k h)^{2}s_{0}^{2}c_{0}^{2}}{2}[J_{m-1}(2khc_{0})+J_{m+1}(2khc_{0})-
J_{m+3}(2khc_{0})-J_{m-3}(2khc_{0})]\Big]~,
\end{multline}
the consequence of which is
\begin{multline}\label{ray_470}
    _{2}B_{m}=-(-\imath)^{m}J_{m}(2khc_{0})+
    \epsilon(kh)^{2}\frac{s_{0}}{2}(-\imath)^{m}
    [J_{m-2}(2khc_{0})-J_{m+2}(2khc_{0})]-\\
    \epsilon^{2}\frac{k h}{4c_{0}^{3}}(-i)^{m}\Big[[J_{m+1}(2khc_{0})-J_{m-1}(2khc_{0})]+\\
    m[J_{m-1}(2khc_{0})+J_{m+1}(2khc_{0})]
    -mk h c_{0}[J_{m-2}(2khc_{0})-J_{m+2}(2khc_{0})]-\\
m\frac{(k h)^{2}s_{0}^{2}c_{0}^{2}}{2}[J_{m-1}(2khc_{0})+J_{m+1}(2khc_{0})-
J_{m+3}(2khc_{0})-J_{m-3}(2khc_{0})]\Big]~.
\end{multline}
\section{The $l=0,1,2$ $\epsilon$-perturbation solutions for a normallly-incident plane wave
striking a sinusoidal boundary}
Now, I give $_{0}B_{m}$, $_{1}B_{m}$, $_{2}B_{m}$ for $\theta^{i}=0$ (i.e., for
$s_{0}=0~,~c_{0}=1$) in order to compare my $\epsilon$-perturbation solutions to those of
Rayleigh.
\subsection{$l=0$}
From (\ref{ray_420}) I find
\begin{equation}\label{ray_480}
    _{0}B_{0}=-J_{0}(2kh)~,~ _{0}B_{1}=\imath J_{1}(2kh)~,~_{0}B_{2}=J_{2}(2kh)~,~
\end{equation}
whereas Rayleigh found (see (\ref{ray_20a})-(\ref{ray_20c}))
\begin{equation}\label{ray_490}
    _{0}A_{0}=-J_{0}(2kh)~,~ _{0}A_{1}=-2\imath J_{1}(2kh)~,~_{0}A_{2}=2J_{2}(2kh)~,~
\end{equation}
which are the same for $m=0,2$ since, I should have found, $A_{0}=B_{0}~,~~A_{2}=2B_{2}$, but
different (by sign) for $m=1$ since, I should have found, $A_{1}=2B_{1}$.
\subsection{$l=1$}
From (\ref{ray_440}) I find
\begin{equation}\label{ray_500}
    _{1}B_{0}=-J_{0}(2kh)~,~ _{1}B_{1}=\imath J_{1}(2kh)~,~_{1}B_{2}=J_{2}(2kh)~,~
\end{equation}
whereas Rayleigh found (see (\ref{ray_20a})-(\ref{ray_20c}))
\begin{equation}\label{ray_510}
    _{1}A_{0}=-J_{0}(2kh)~,~ _{1}A_{1}=-2\imath J_{1}(2kh)~,~_{1}A_{2}=2J_{2}(2kh)~,~
\end{equation}
which, again, are the same for $m=0,2$ since, I should have found, $A_{0}=B_{0}~,~~A_{2}=2B_{2}$, but different (by sign) for $m=1$ since, I should have found, $A_{1}=2B_{1}$.
\subsection{$l=2$}
From (\ref{ray_450}) I find
\begin{multline}\label{ray_520}
    _{2}B_{0}=-J_{0}(2kh)-\epsilon^{2}\frac{k h}{2}J_{1}(2kh)~,~
    _{2}B_{1}=\imath J_{1}(2kh)+\epsilon^{2}\imath k h J_{2}(k h)~,\\
    ~_{2}B_{2}=J_{2}(2kh)-\epsilon^{2}\frac{k h}{4}\Big(J_{1}(2kh)+3J_{3}(2kh)-2kh[J_{0}(2kh)-J_{4}(2kh)]\Big)~,~
\end{multline}
whereas Rayleigh found (see (\ref{ray_20a})-(\ref{ray_20c}))
\begin{equation}\label{ray_530}
    _{2}A_{0}=-J_{0}(2kh)-\epsilon^{2}\frac{kh}{2}J_{1}(2kh)~,~
    _{2}A_{1}=-2\imath J_{1}(2kh)-\epsilon^{2}\imath\frac{kh}{2}J_{2}(2kh)~,~_{1}A_{2}=2J_{2}(2kh)~,~
\end{equation}
which, again,  are the same for $m=0,2$ since, I should have found, $A_{0}=B_{0}~,~~A_{2}=2B_{2}$, but different (by sign) for $m=1$ since, I should have found, $A_{1}=2B_{1}$.
\subsection{Comments}\label{Commens}
I have thus shown, by comparisons with my mathematically-consistent perturbation solutions,
that Rayleigh's rather cursory perturbation scheme (for the only case he treated in which
$\theta^{i}=0~,~f(x)=h\cos\frac{2\pi}{d}x~,~\text {perturbation orders}~0,1,2$) appears to lead to zeroth, first, and
second diffraction-order solutions that are only correct for $A_{0}$. Of course, this statement
is sound only if more evidence is given that my perturbation solutions are really correct.
This will be done in the next section.
\section{Demonstration that the perturbation solutions satisfy the Conservation of Flux
Relation}\label{CFR}
In pages 94-95 of Rayleigh's 'Theory of Sound', he demonstrates (in my notation) that the
scattering amplitudes $B_{m}$ (for a plane wave of unit amplitude striking an impervious,
periodic boundary) must satisfy the so-called 'Conservation of Flux Relation' (C F R) whose
expression, for arbitrary $\theta^{i}$ and 'arbitrary' $f(x)$, is
\begin{equation}\label{ray_540}
    \mathfrak{F}=\sum_{n=-\infty}^{\infty}\mathfrak{F}_{n}=
    \sum_{n=-\infty}^{\infty}\|B_{n}\|^{2}\Re \frac{c_{n}}{c_{0}}=1~,
\end{equation}
wherein $\mathfrak{F}$ designates the total normalized flux and
$\mathfrak{F}_{n}=\|B_{n}\|^{2}\Re \frac{c_{n}}{c_{0}}$ the $n$-t h order partial normalized
flux (oft-named '$n$-t h order efficiency' in the optics community).

Note that Rayleigh did not attempt to show that his high-frequency perturbation solutions satisfy his flux-conservation relation. I do this in the following subsection.
\subsection{Application of the perturbation method in the C F R}\label{CFRpert}
 I proceed as previously by expanding $B_{n}$, $B_{n}^{*}$, $\Re \frac{c_{n}}{c_{0}}$,
$\mathfrak{F}_{n}$ and $\mathfrak{F}$ in the Taylor series:
\begin{equation}\label{ray_550}
        B_{n}=\sum_{p=0}^{\infty}B_{n}^{(p)}\epsilon^{p}~,~
    B_{n}^{*}=\sum_{l=0}^{\infty}B_{n}^{*(l)}\epsilon^{l}~,~
    \Re\frac{c_{n}}{c_{0}}=\sum_{m=0}^{\infty}\Big(\Re\frac{c_{n}}{c_{0}}\Big)^{(m)}\epsilon^{m}~,~ \mathfrak{F}_{n}=\sum_{\nu=0}^{\infty}\mathfrak{F}_{n}^{\nu}\varepsilon^{\nu}~,~
    \mathfrak{F}=\sum_{j=0}^{\infty}\mathfrak{F}^{(j}\epsilon^{j}~,
\end{equation}
in which:
\begin{equation}\label{ray_560}
B_{n}^{(p)}=\frac{1}{p!}\frac{\partial^{p}}{\partial\epsilon^{p}}B_{n}\Big|_{\epsilon=0}~,~
B_{n}^{*(l)}=\frac{1}{l!}\frac{\partial^{l}}{\partial\epsilon^{l}}B_{n}^{*}\Big|_{\epsilon=0}~,~
\Big(\Re\frac{c_{n}}{c_{0}}\Big)^{(m)}=\frac{1}{m!}\frac{\partial^{m}}{\partial\epsilon^{m}}
\Big(\Re\frac{c_{n}}{c_{0}}\Big)\Big|_{\epsilon=0}~,~
\end{equation}
so that
\begin{equation}\label{ray_570}
\mathfrak{F}_{n}^{(\nu)}=\sum_{p=0}^{\nu}\sum_{l=0}^{\nu-p}B_{n}^{(p)}B_{n}^{*(l)}
\Big(\Re\frac{c_{n}}{c_{0}}\Big)^{(\nu-p-l)}~;~\nu=0,1,2...~.
\end{equation}
\subsection{The $\nu=0$ perturbation order}
Eq. (\ref{ray_570}) yields
\begin{equation}\label{ray_580}
\mathfrak{F}_{n}^{(0)}=B_{n}^{(0)}B_{n}^{*(0)}
\Big(\Re\frac{c_{n}}{c_{0}}\Big)^{(0)}~.
\end{equation}
It is readily shown that $B_{n}^{*(0)}=(B_{n}^{(0)})^{*}$ whereas $\Big(\Re\frac{c_{n}}{c_{0}}\Big)^{(0)}=1$, so that
\begin{equation}\label{ray_590}
\mathfrak{F}_{n}^{(0)}=\|B_{n}^{(0)}\|^{2}~,
\end{equation}
or
\begin{equation}\label{ray_600}
\mathfrak{F}_{n}^{(0)}=\int_{0}^{1}d\zeta'\exp[-\imath(2\pi n\zeta'+2c_{0}kg')]
\int_{0}^{1}d\zeta\exp[\imath(2\pi n\zeta+2c_{0}kg)]~.
\end{equation}
Next, consider the zeroth-order total normalized flux $\mathfrak{F}^{(0)}=\sum_{n=-\infty}^{\infty}\mathfrak{F}_{n}^{(0)}$:
\begin{equation}\label{ray_610}
\mathfrak{F}^{(0)}=\int_{0}^{1}d\zeta'\exp[-\imath 2c_{0}kg')]
\int_{0}^{1}d\zeta\exp[\imath 2c_{0}kg)]\sum_{n=-\infty}^{\infty}\exp[-\imath(2\pi n(\zeta'-\zeta)]~,
\end{equation}
which becomes, on account of (\ref{ray_310})
\begin{equation}\label{ray_620}
\mathfrak{F}^{(0)}=\int_{0}^{1}d\zeta'\exp[-\imath 2c_{0}kg')]
\int_{0}^{1}d\zeta\exp[\imath 2c_{0}kg)]\delta(\zeta'-\zeta)=\int_{0}^{1}d\zeta'=1~
\end{equation}

Thus, $_{0}{\mathfrak{F}}=\mathfrak{F}^{(0)}=1$, which shows that the zeroth-order perturbation solution satisfies, as it should, the CFR. This also supports the contention that this solution is correct (for 'arbitrary' $f$ and $\theta^{i}$.
\subsection{The $\nu=1$ perturbation order}
Eq. (\ref{ray_570}) yields
\begin{equation}\label{ray_630}
\mathfrak{F}_{n}^{(1)}=B_{n}^{(0)}B_{n}^{*(0)}
\Big(\Re\frac{c_{n}}{c_{0}}\Big)^{(1)}+B_{n}^{(0)}B_{n}^{*(1)}
\Big(\Re\frac{c_{n}}{c_{0}}\Big)^{(0)}+B_{n}^{(1)}B_{n}^{*(0)}
\Big(\Re\frac{c_{n}}{c_{0}}\Big)^{(0)}.
\end{equation}
It is readily shown that $B_{n}^{*(1)}=(B_{n}^{(1)})^{*}$ whereas $c_{n}^{(1}/c_{0} =-ns/c_{0}$, so that the first-order CFR becomes
\begin{equation}\label{ray_640}
\mathfrak{F}_{n}^{(1)}=\|B_{n}^{(0)}\|^{2}\Big(\frac{-ns_{0}}{c_{0}}\Big)+
2\Re\Big[B_{n}^{(0)}B_{n}^{1*}\Big]=\mathfrak{F}_{n}^{(0)}\Big(\frac{-ns_{0}}{c_{0}}\Big)+
2\Re\Big[B_{n}^{(0)}B_{n}^{1*}\Big]~,
\end{equation}
or
\begin{multline}\label{ray_650}
\mathfrak{F}_{n}^{(1)}=\\
\mathfrak{F}_{n}^{(0)}\Big(\frac{-n s_{0}}{c_{0}}\Big)+
\Re\Big[\Big(\frac{-\imath 2\pi k^{2}s_{0}}{\pi}\Big)\int_{0}^{1}d\zeta'\exp[-\imath(2\pi n\zeta'+2c_{0}kg')]
\int_{0}^{1}d\zeta g g_{,\zeta}\exp[\imath(2\pi n\zeta+2c_{0}kg)]\Big]~.
\end{multline}
Next, consider the first-order total normalized flux $\mathfrak{F}^{(1)}=\sum_{n=-\infty}^{\infty}\mathfrak{F}_{n}^{(1)}=
\mathfrak{F}_{n}^{(1)1}+\mathfrak{F}_{n}^{(1)2}$:
\begin{equation}\label{ray_660}
\mathfrak{F}^{(1)1}=\Big(\frac{-s_{0}}{c_{0}}\Big)\int_{0}^{1}d\zeta'\exp[-\imath 2c_{0}kg')]
\int_{0}^{1}d\zeta\exp[\imath 2c_{0}kg)]\sum_{n=-\infty}^{\infty}n\exp[-\imath(2\pi n(\zeta'-\zeta)]~,
\end{equation}
which becomes, on account of (\ref{ray_320})
\begin{equation}\label{ray_670}
\mathfrak{F}^{(1)1}=\Big(\frac{-s_{0}}{c_{0}}\Big)\Big(\frac{-\imath 2kc_{0}}{2\pi}\Big)
\int_{0}^{1}d\zeta'g'_{,\zeta'}=0~,
\end{equation}
the vanishing nature of the latter integral being due to the periodicity of $g$.
On the other hand,
\begin{multline}\label{ray_680}
\mathfrak{F}^{(1)2}=\sum_{n=-\infty}^{\infty}
\Re\Big[\Big(\frac{-\imath 2\pi k^{2}s_{0}}{\pi}\Big)\int_{0}^{1}d\zeta'\exp[-\imath(2\pi n\zeta'+2c_{0}kg')]
\int_{0}^{1}d\zeta g g_{,\zeta}\exp[\imath(2\pi n\zeta+2c_{0}kg)]\Big]=\\
\Re\Big[\Big(\frac{-\imath 2\pi k^{2}s_{0}}{\pi}\Big)\int_{0}^{1}d\zeta'\exp[-\imath 2c_{0}kg')]
\int_{0}^{1}d\zeta g g_{,\zeta}\exp[\imath 2c_{0}kg)]\delta(\zeta'-\zeta\Big]=\\
\Re\Big[\Big(\frac{-\imath 2\pi k^{2}s_{0}}{\pi}\Big)
\int_{0}^{1}d\zeta'g'g'_{,\zeta'}\Big]=
\Re\Big[\Big(\frac{-\imath 2\pi k^{2}s_{0}}{\pi}\Big)g'^{2}\Big|_{z'=0}^{1} \Big]=0
~,
\end{multline}
the result on the last line of this equation being due to the periodicity of $g$.
It ensues that $\mathfrak{F}^{(1)}=0$.

Thus, $_{1}{\mathfrak{F}}=\mathfrak{F}^{(0)}+\varepsilon \mathfrak{F}^{(1)}=1$, which shows that the first-order perturbation solution satisfies, as it should, the C F R. This also supports the contention that this solution is correct (for 'arbitrary' $f$ and $\theta^{i}$.
\subsection{The $\nu\ge 2$ perturbation orders}
I have not pursued these demonstrations for the $\nu\ge 2$ perturbation orders because of the algebraic complexity this would entail.
\section{Conclusion}
 The unknowns in Rayleigh's seemingly-rigorous linear system of equations are the plane-wave reflection coefficients. Although the latter can easily be found  numerically for low temporal frequencies and/or modest grating amplitude-to-grating period ratios \cite{wi80} it becomes difficult, or nearly impossible, to achieve this numerical determination, for medium-to-high temporal frequencies, as is sometimes the case, notably in optics, when a characteristic dimension of the scatterer can be much larger than the wavelength. Be this as it may, the optics community has always massively relied on high-frequency heuristic paradigms (even to the point of neglecting the vectorial nature of the electromagnetic wavefield) to predict the physical phenomena with which it was confronted, usually at the expense of ignoring a mathematical approach to solving the boundary-value problems at the root of Maxwell's equations. This is why Rayleigh's plane-wave representation of the wavefield, together with his high-frequency perturbation method for the determination of the plane-wave amplitudes, mark an authentic rupture in the way people in the optics (as well as acoustics) communitie(s) approach their scattering problems \cite{bs63}. Unfortunately, Rayleigh's non-heuristic methods for the diffraction grating were published in a rather restricted setting: normal plane-wave incidence and the sole sinusoidal grating profile. Moreover, Rayleigh made no significant attempt to relate his perturbation solutions to the heuristic solutions that existed before his own. The same can be said concerning the relation of the more-recent heuristic solutions \cite{bs63},\cite{fo69} to the mathematically-sound  high-frequency solution (shown herein to satisfy what resembles a conservation of energy relation), which is the reason why I propose to do this, for arbitrary grating profiles and incident angles, in a forthcoming publication. In another forthcoming publication, I will also numerically compare the various-order perturbation predictions to those that stem from the full-blown numerical resolution of Rayleigh's linear system of equations, for three types of grating profiles.

\end{document}